\documentclass[journal,10pt]{IEEEtran}
\usepackage{amssymb}
\usepackage{amsmath}
\usepackage{cite}
\usepackage{url}
\usepackage{xcolor}
\usepackage{cite,graphicx,amsmath,amssymb}
\usepackage{subfigure}
\usepackage{fancyhdr}
\usepackage{mdwmath}
\usepackage{mdwtab}
\usepackage{caption}
\usepackage{amsthm}
\usepackage{setspace}
\usepackage{algorithm}
\usepackage{algorithmic}
\usepackage{makecell}
\usepackage{diagbox}
\usepackage{multirow}

\usepackage{adjustbox}
\usepackage{blindtext}

\usepackage{subfigure}
\usepackage{float}

\newtheorem{theorem}{Theorem}

\newtheorem{lemma}{Lemma}

\newtheorem{corollary}{Corollary}

\allowdisplaybreaks
\makeatletter
\def\ScaleIfNeeded{%
\ifdim\Gin@nat@width>\linewidth \linewidth \else \Gin@nat@width
\fi } \makeatother

\begin{document}

\title{Two-Stage Hierarchical Beam Training for Near-Field Communications}

\author{
	Chenyu~Wu, Changsheng~You, \IEEEmembership{Member, IEEE}, 
	Yuanwei~Liu, \IEEEmembership{Senior Member,~IEEE}, 
	\\Li~Chen, \IEEEmembership{Senior Member,~IEEE},
	and Shuo Shi, \IEEEmembership{Member,~IEEE} 	\vspace{-2em}
		\thanks{C. Wu is with the School of Electronic and Information Engineering, Harbin Institute of Technology (HIT), Harbin 150001, China, and also with the Department of Electrical and Electronic Engineering, Southern University of Science and Technology (SUSTech), Shenzhen 518055, China (e-mail: wuchenyu@hit.edu.cn).}
		\thanks{C. You is with the Department of Electrical and Electronic Engineering, Southern University of Science and Technology (SUSTech), Shenzhen 518055, China (email: youcs@sustech.edu.cn). 
		}
		\thanks{Y. Liu is with the School of Electronic Engineering and Computer Science, Queen Mary University of London, London E1 4NS, UK (email: yuanwei.liu@qmul.ac.uk).}
		\thanks{L. Chen is with the CAS Key Laboratory of Wireless-Optical Communications, University of Science and Technology of China, Hefei 230027, China (e-mail: chenli87@ustc.edu.cn).}
		\thanks{S. Shi is with  the School of Electronic and Information Engineering, Harbin Institute of Technology (HIT), Harbin 150001, China (e-mail: crcss@hit.edu.cn).} \thanks{\textit{(Corresponding author: Changsheng You)}}	
	}
\maketitle
\begin{abstract}
Extremely large-scale array (XL-array) has emerged as a promising technology to improve the spectrum efficiency and spatial resolution of future wireless systems. However, the huge number of antennas renders the users more likely to locate in the \textit{near-field} (instead of the far-field) region of the XL-array with spherical wavefront propagation. This inevitably incurs prohibitively high beam training overhead since it requires a two-dimensional (2D) beam search over both the angular and distance domains. To address this issue, we propose in this paper an efficient \textit{two-stage hierarchical beam training} method for near-field communications. Specifically, in the first stage, we employ the central sub-array of the XL-array to search for a coarse user direction in the angular domain with conventional far-field hierarchical codebook. Then, in the second stage, given the coarse user direction, we progressively search for the fine-grained user direction-and-distance in the polar domain with a dedicatedly designed codebook. Numerical results show that our proposed two-stage hierarchical beam training method can achieve over 99\% training overhead reduction as compared to the 2D exhaustive search, yet achieving comparable rate performance.

\end{abstract}


%
%
%
%
\section{Introduction}
Extremely large-scale array/surface (XL-array/surface) has emerged as a promising technology to achieve ultra-high spectrum efficiency and super spatial resolution in future sixth-generation (6G) communication systems~\cite{nf_mag,zr_survey,ris_survey2,survey_ris,zy,xl,SRAR}. Specifically, with wireless communications migrating to higher frequency bands, such as millimeter-wave (mmWave) and even terahertz (THz)~\cite{thz}, XL-array/surface is expected to be widely deployed at the base station (BS) to compensate for the severe path-loss via highly directional beamforming.

With the sharp increase in the carrier frequency and number of antennas, the well-known \textit{Rayleigh distance} will expand to dozens to hundreds of  meters, which means that the users are more likely to locate in the \textit{near-field} region of the XL-array~\cite{nf_mag}. Different from conventional \textit{far-field} communications with its electromagnetic (EM) field simply modeled as planar waves, the EM field of near-field communications should be accurately modeled by \textit{spherical} waves. This unique channel characteristic can be delicately leveraged to enable flexible \textit{beam focusing}, for which the beam energy can be focused on a specific spatial location rather than a specific direction as in conventional far-field communications~\cite{beamfocusing}. Despite many advantages, near-field communications also face critical challenges in practical implementation, such as low-overhead beam training. In this paper, we propose an efficient two-stage hierarchical beam training method to significantly reduce the training overhead of existing schemes.


\subsection{Related Works}
\textit{1) Near-field Wireless Communications:} Near-field communications have attracted extensive research attention in recent years that exploits XL-arrays/surfaces for improving the performance of wireless systems. For example, a new concept of location division multiple access (LDMA) was proposed in \cite{wu2022multiple} to enhance the multiple accessibility and system capacity in the near-field region. Particularly, the additional spatial resolution of near-field communications was exploited by the XL-array to serve multiple users simultaneously without introducing severe inter-user interference, even when the users are located in the same spatial directions. In addition, the near-field beam focusing property was utilized to improve the charging efficiency of wireless power transfer (WPT) systems\cite{WPT}. 
In \cite{beixiong1}, an XL-surface aided communication system  was studied, where the XL-surface is integrated into an access point to achieve transmit diversity and passive beamforming simultaneously.
For near-field localization, the authors in \cite{hcw1} optimized the XL-array phase-shifts to improve the localization accuracy, while the Cram\'{e}r-Rao bound (CRB) performance was studied in\cite{chenli}. Moreover, the coexistence of far-field and near-field communications may also appear in future wireless communication systems\cite{crossfield}. In this context, a mixed near- and far-field communication scenario was investigated in \cite{yunpu2}, where the inter-user interference at the near-field user caused by the far-field beam was analytically characterized.


\textit{2) Conventional Far-field Beam Training:} To fully reap the prominent beamforming gain brought by XL-array/surface, it is indispensable to perform efficient beam training at the BS to establish a high-quality initial link before implementing channel estimation and data transmission~\cite{wang2022beam}.
For far-field beam training, the BS conducts beam sweeping through multiple directional beams predefined in a codebook, while each user identifies its optimal beam (codeword) and feed its index back to the BS\cite{6847111}. There have been substantial works on the far-field beam training (codebook) design with the aim to improve beam training accuracy and/or reduce its overhead, see e.g., \cite{bisection,xiao,csyou1,qch}. Specifically, in \cite{bisection}, the authors proposed a hierarchical beam training method, where wide beams are generated by activating a sub-array of the full-array. The beam training accuracy was further improved in \cite{xiao} by designing another hierarchical codebook that jointly utilizes the sub-array and deactivation techniques. Moreover, a novel multi-beam training method was proposed in \cite{csyou1} that steers multiple beams simultaneously at each time for reducing multi-user beam training overhead. In \cite{qch}, the authors  proposed a simultaneous multi-user hierarchical beam training scheme for mmWave massive multi-input multi-output (MIMO) systems, wherein an alternative minimization method was proposed to design the hierarchical codebook, based on which the beam training scheme at the BS was adaptively adjusted according to the beam training results of previous layers.

However, these far-field beam training methods may not be applicable to the near-field case, since the near-field user channel is related to not only the spatial angle-of-departure/arrival (AoD/AoA), but also the (effective) BS-user distance. In particular, when the far-field beams are applied to near-field beam training, there may appear the so-called \textit{energy-spread} effect \cite{nf_exhaustive}, in which the energy of a beam steering at a specific direction will be spread into multiple angles. As such, directly applying the far-field beam training methods \cite{bisection,xiao,csyou1,qch} to the near-field scenario will result in severe performance degradation.


\textit{3) Near-field Beam Training:} To tackle this issue, several research efforts have been made to design efficient near-field beam training methods. Specifically, the authors in \cite{nf_exhaustive} proposed a new polar-domain codebook, wherein the angular domain is uniformly sampled while the distance domain is non-uniformly sampled. Based on this polar-domain codebook, the exhaustive-search based beam training method will incur prohibitively high training overhead, which is the product of the number of BS antennas and sampled distances. To reduce the training overhead, an efficient two-phase beam training method was recently proposed in \cite{two_phase}, which decomposes the joint estimation of the spatial angle and distance into two separate phases. This scheme exploits a key observation that when far-field beams are used for beam training, the true user spatial direction approximately locates in the middle of a dominant angular region with high received signal powers. In \cite{dl}, the authors  proposed a deep-learning based beam training scheme which predicts the best near-field codeword by using neural networks with measurements of the beam gains of conventional far-field wide beams. For near-field wideband communications,
the effect of beam split was utilized in \cite{wideband} to enable fast beam training by focusing beams at different frequencies on different desired locations. Besides, the authors in \cite{wei} proposed a near-field hierarchical beam training method 
based on a new rectangular coordinate system. However, this scheme is highly sensitive to the sampling steps and relies on the prior knowledge of the candidate region. In \cite{dai_hier}, the authors 
customized a near-field hierarchical beam training scheme for the digital precoding architecture by designing a near-field multi-resolution codebooks using the Gerchberg–Saxton (GS) based algorithm. 


\subsection{Our Contributions}

In this paper, we propose an efficient near-field hierarchical beam training design to further reduce the training overhead of existing near-field beam training methods. The main contributions of this paper are summarized as follows.
\begin{itemize}
	\item First, we present one key observation that by properly activating a central sub-array of the XL-array for beam training, the near-field energy-spread effect can be effectively mitigated. This thus inspires us to design a simple yet efficient \textit{two-stage hierarchical} beam training method. Specifically, in the first stage, we estimate the \textit{coarse user direction} by employing the sub-array technique to implement the conventional far-field hierarchical beam training. Then, in the second stage, we further progressively resolve \textit{the user direction and its effective distance} by devising a dedicated polar-domain hierarchical codebook as well as a new beam training method.
	\item  Second, we show that the training overhead of the proposed two-stage hierarchical beam training design scales with the number of XL-array antennas, $N$, in the order of $\mathcal{O}(\log_2(N))$. Moreover, we discuss the extensions of the proposed beam training design to other systems with planar arrays and metasurfaces. 
	\item Finally, numerical results are provided to verify the effectiveness of our proposed two-stage hierarchical beam training design. In particular, we show that our proposed hierarchical beam training design can achieve over $99\%$ and $95\%$ beam training overhead reduction as compared with the exhaustive-search based \cite{nf_exhaustive} and two-phase \cite{two_phase} near-field beam training schemes, respectively, yet achieving comparable rate performance.	
\end{itemize}

The rest of the paper is organized as follows. In Section \ref{sec2}, we present the near-field communication system model. Several existing benchmark schemes for near-field beam training are introduced in Section \ref{sec3}.
In Section \ref{sec4}, we present our key observations and the proposed two-stage hierarchical beam training method. In Section V, we discuss several key parameters of
our design and the extensions to other systems. Numerical results are shown in Section \ref{sec5}, followed by the conclusions in Section \ref{sec6}. 

\textit{Notations:} For a vector $\mathbf{a}$,  $(\mathbf{a})^T$ and $(\mathbf{a})^H$ denote its transpose and conjugate transpose, respectively. For a set $\mathcal{S}$, $\mathcal{S}[n]$ returns its $n$-th element. 
$\lceil \cdot \rceil$ and $\lfloor \cdot \rfloor$ denote the ceil and floor operator, respectively. $\mathbf{0}_n$ represents a vector with $n$ zero elements. ${\mathcal{C}\mathcal{N}(\mu,\sigma^2)}$ denotes the circularly symmetric complex Gaussian distribution with mean $\mu$ and variance $\sigma^2$. $\mathbb{I}(\cdot)$ is the indicator function. $\otimes$ stands for the Kronecker product.

\section{System Model}\label{sec2}
\label{s2}

\begin{figure}[t]
	\centering
	\includegraphics[width=0.45\textwidth]{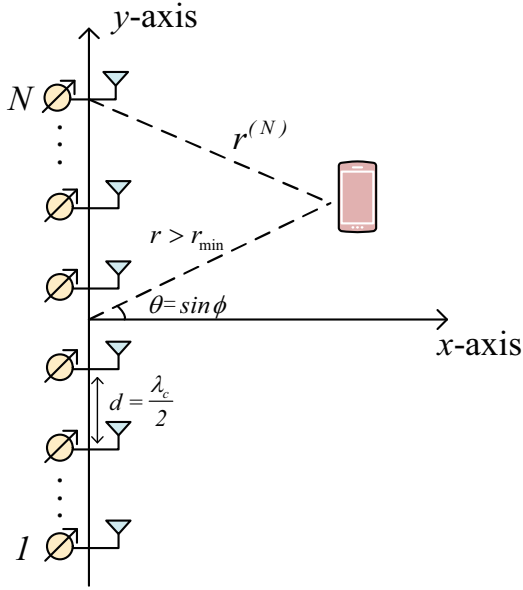}\\
	\caption{A narrow-band XL-array communication system.}\label{model}
\end{figure}

%

As shown in Fig.~\ref{model}, we consider a narrow-band downlink communication system, where a BS equipped with an $N$-antenna uniform linear array (ULA) communicates with a single-antenna user.\footnote{The proposed beam training method can be readily applied to the multi-user scenario by employing time division multiple access (TDMA). Therefore, without loss of generality, we consider a typical user in this paper.} The antennas are placed along the $y$-axis with the coordinate of the $n$-th antenna given
by $(0,\delta_n{d})$, where $\delta_n=\frac{2n-N-1}{2}$, and $d = \frac{\lambda_c}{2}$ is the antenna spacing with $\lambda_c$ denoting the carrier wavelength. Then, based on the
spherical wavefront propagation model, the near-field line-of-sight (LoS) channel from the BS to the user can be modeled as~\cite{nf_exhaustive}
\begin{equation}\label{nf_channel}
	\mathbf{h}=\sqrt{N}h\mathbf{b}(\theta,r),
\end{equation}
where $h=\frac{\sqrt{{\rho_0}}}{r}e^{-\jmath{\frac{2\pi}{\lambda_c}}r}$ is  
the complex channel gain with $\rho_0$ denoting the reference channel power gain at a distance of 1 m, $\theta$ is the sine of the spatial AoD $\phi$, and $r$ denotes the distance between the geometric center of the BS and the user. Besides, $\mathbf{b}(\theta,r)$ denotes the near-field steering vector, given by\footnote{In this paper, the near-field steering vector can represent the response of a ULA/sub-array with arbitrary numbers of antennas.}
\begin{equation}
	\mathbf{b}(\theta,r)=\frac{1}{\sqrt{N}}\left[ e^{-\jmath{\frac{2\pi}{\lambda_c}}(r^{(1)}-r)}, ..., e^{-\jmath{\frac{2\pi}{\lambda_c}}(r^{(N)}-r)}\right]^T, 
\end{equation}
where $r^{(n)} =\sqrt{r^2(1-\theta^2)+(r\theta-\delta_n{d})^2}$ represents the distance between the $n$-th antenna at the BS and the user. Note that we assume in this paper that the user is located in the Fresnel region, i.e., $r>r_{\min} =\max\{\frac{1}{2}\sqrt{\frac{D^3}{\lambda_c}},1.2D\}$, where the amplitude variations across the antennas are negligible \cite{ff_mag,9723331}. It is worth pointing out that when $r$ is sufficiently large, $\mathbf{b}(\theta,r)$ is equivalent to the conventional far-field steering vector 
\begin{equation}\label{ff_steering}
	\mathbf{a}(\theta)=\frac{1}{\sqrt{N}}\left[ 1, e^{-\jmath \pi \theta}, ..., e^{-\jmath \pi(N-1)\theta}\right]^T. 
\end{equation}
This is because $r^{(n)}\approx{r-\delta_nd\theta}$ when $r$ is sufficiently large. Thus, the near-field steering vector is a general model that characterizes the channel accurately.

Based on the BS-user channel in \eqref{nf_channel}, the received signal at the user is given by
\begin{equation}\label{rpower}
	y(\mathbf{v})=\mathbf{h}^H\mathbf{v}x+z_0,
\end{equation}
where $x$ denotes the transmitted symbol of the BS, $\mathbf{v}$ represents the analog phase-shifter (PS) based transmit beamforming vector at the BS, and $z_0\sim{\mathcal{C}\mathcal{N}(0,\sigma^2)}$ is the additive white Gaussian noise (AWGN) at the user with power $\sigma^2$.

From \eqref{rpower}, it can be easily obtained that the optimal BS beamforming vector is $\mathbf{v}^*=\mathbf{b}(\theta,r)$. In practice, the beam training method is usually performed to establish a high-quality initial link before channel estimation and data transmission, by finding the optimal beam codeword that matches $\mathbf{v}^*$. 

\section{Benchmark Schemes}\label{sec3}
In this section, we introduce existing beam training schemes for near-field communications and point out their limitations.

\subsection{Exhaustive-Search Based Near-field Beam Training}\label{3.1}

Different from the far-field beam training, the near-field beam training requires the beam search in both the angular and distance domains. To this end, a polar-domain codebook was proposed in \cite{nf_exhaustive}, which is given by
\begin{equation}\label{ex_codebook}
	\mathbf{W}^{\rm{pol}}=\left\{\mathbf{w}^{\rm pol}_{n,s}=\mathbf{b}(\theta_n,r_n^{(s)})|n\in\mathcal{N},s\in\mathcal{S}\right\},
\end{equation}
where $\mathcal{N}=\{1,...,N\}$, $\mathcal{S}=\{1,...,S\}$, $\theta_n=\frac{2n-N-1}{N}$ represents the pointing angle of each codeword in $\mathbf{W}^{\rm pol}$, which is uniformly sampled in the spatial angular domain $[-1,1]$, $S$ and $r_n^{(s)}$ denote the number of sampled distances and the $s$-th sampled distance for each angle $\theta_n$, respectively. Specifically, $r_n^{(s)}$ is given by
\begin{equation}\label{5}
	r_n^{\left(s\right)}=\frac{1}{s}S_\Delta(1-(\theta_n)^2),s=0,1,...,S-1,
\end{equation}
where $S_\Delta$ is a constant~\cite{nf_exhaustive}. Note that $\forall n$, the distance intervals, $r_n^{(s+1)}-r_n^{(s)}, s=0,...,S-1$, is larger as the distance increases since the near-field effect gradually diminishes with distance.

Based on the polar-domain codebook in \eqref{ex_codebook}, we denote
\begin{equation}\label{opt_codeword}
	(n^*,s^*)=\arg \max_{(n,s)}(|\mathbf{b}^H(\theta, r) \mathbf{w}^{\rm pol}_{n,s}|^2  )
\end{equation}
as the optimal codeword index for the user. In practice, 
the exhaustive-search based method can be applied to find the best beam that attains the maximum received signal-to-noise ratio (SNR) at the user. However, this method incurs a total overhead of $NS$,
which is unaffordable when $N$ is large. For example, when $N=512, S=6$, the total beam training overhead is 3072.

\subsection{Two-phase Near-field Beam Training}\label{3.2}

To reduce the huge beam training overhead of the exhaustive-search based scheme, a fast near-field beam training method was proposed in \cite{two_phase} that separately estimates the spatial user direction and distance in two sequential phases. Specifically, a key observation shows that when the far-field beams are used for beam training, the true spatial direction of the near-field user approximately lies in the middle of the dominant angular region with sufficiently high beam powers (as illustrated in the blue curve of Fig.~\ref{fig2}). Given this observation, in the first phase, we employ the conventional angular-domain far-field codebook given by 
\begin{equation}\label{ff_codebook}
	\mathbf{W}^{\rm ang}=\left\{\mathbf{w}^{\rm ang}_{n}=\mathbf{a}(\theta_n)\Big| n=1,...,N, \theta_n=\frac{2n-N-1}{N}\right\}
\end{equation}
to perform beam sweeping over the whole angular domain. Then, the $K$ directions that lie in the middle of the dominant angular region are selected as candidate spatial angles. In the second phase, the effective distance of the user is estimated by searching over the distance domain according to \eqref{5} for each candidate spatial angle. Compared with the exhaustive-search based beam training method,  
the beam training overhead of this method is reduced to $N+KS$. However, this overhead is still proportion to $N$, which is prohibitively high when $N$ is large.  For instance, when $N=512, S=6,K=1$, the total overhead of the two-phase scheme in \cite{two_phase} is 518.


\subsection{Conventional Far-field Hierarchical Beam training}
Consider the conventional far-field \textit{hierarchical beam training method} and its issues when applied in the near-field beam training.
The basic idea of hierarchical beam training is to first
search for a coarse user direction by employing wide beams and then progressively identify fine-grained user spatial directions by using narrower beams. 
The corresponding far-field hierarchical codebook can be implemented by either the deactivation or the joint sub-array and deactivation techniques, with the details given in \cite{bisection} and \cite{xiao}, respectively. When $N=512$, the total overhead of far-field hierarchical beam training is $2\log_2(N)=18$. 

However, the conventional far-field hierarchical beam training is no longer applicable to the near-field case due to the unique near-field \textit{energy-spread} effect \cite{nf_exhaustive}, which will be specified in the next section.

\section{Proposed Two-stage Hierarchical Beam Training}\label{sec4}
In this section, we propose a new two-stage hierarchical beam training method to significantly reduce the overhead of the existing near-field beam training methods, yet achieving superior rate performance.

\subsection{Key Idea}
First, we define the beam gain as follows.

\vspace{0.4em}
\noindent\underline{\textbf{Definition 1.}} \textit{Define $g(\mathbf{b}(\theta,r),\mathbf{w})$ as the normalized beam gain of the beamforming vector $\mathbf{w}$ along the channel steering vector $\mathbf{b}(\theta,r)$, which is given by}
	\begin{equation}
		g(\mathbf{b}(\theta,r),\mathbf{w})=|\mathbf{b}^H(\theta,r)\mathbf{w}|.
	\end{equation}

\begin{figure}[t]
	\centering
	\includegraphics[width=0.5\textwidth]{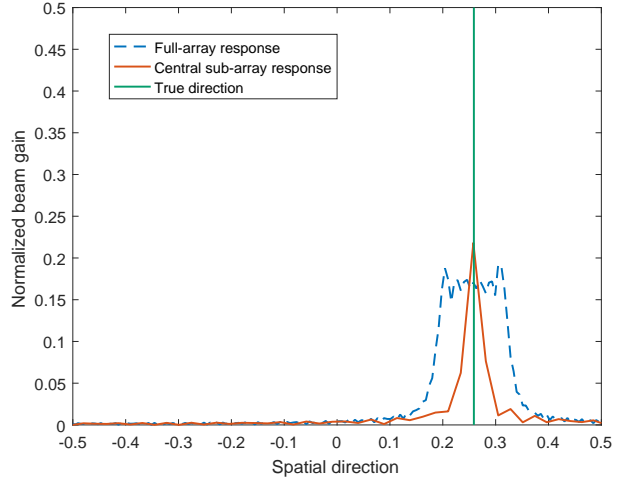}\\
	\caption{Normalized beam gain of the full-array and a central sub-array, respectively. The Xl-array has $N=512$ antennas and the central sub-array includes $N_c=128$ antennas. The carrier frequency $f$ is 100 GHz and the BS-user distance is $r=5$ m.}\label{fig2}
\end{figure}

Next, consider a near-field user with a channel steering vector $\mathbf{b}(\theta,r)$. We present two key observations when the far-field beams based on the full- or sub-array are applied for beam sweeping.

\vspace{0.4em}
\noindent\underline{\textbf{Observation 1}} \textit{(Near-field energy-spread effect)}.  \textit{Consider the full-array beam sweeping with far-field beams given in \eqref{ff_codebook}. The beam gain defined in (8) is large and fluctuates over a \textbf{wide} spatial angular region (see the blue curve of Fig. \ref{fig2}).}

%

This is the so-called \textit{energy-spread} effect for near-field communications~\cite{nf_exhaustive}, which can be explained as follows:  The XL-array can be equally divided into multiple sub-arrays, 
such that the user locates in the far-field region of each sub-array, while it lies in the near-field region of the full-array. This thus leads to different AoDs (or called pointing directions) from each individual sub-array to the user. As such, when the full-array far-field beams are applied for beam sweeping in the neighborhood of the true direction $\theta$, the beam angle will align with the pointing directions of certain sub-arrays, hence yielding a wide angular region with strong beam gains. Therefore, it is expected that if we use the full XL-array for hierarchical beam training, the users cannot distinguish the optimal beam direction by finding the maximum received signal power among all beams. 

\vspace{0.4em}
\noindent \underline{\textbf{Observation 2}} \textit{(Sub-array beam sweeping without near-field effect)}. \textit{Consider the sub-array beam sweeping with far-field beams where only a central sub-array with $N_c=N/4$ antennas are activated, i.e., $\mathbf{w}_i=\left[ \mathbf{0}^T_{\frac{3N}{8}},\mathbf{a}^T\left( -1+\frac{2i-1}{N_c}\right), \mathbf{0}^T_{\frac{3N}{8}}\right]^T,i=1,...,N_c$. The beam gain of the central sub-array has a sharp peak pointing to the true direction of the user, and the normalized beam gain of the central sub-array approximates that of the full-array (see the red curve in Fig. \ref{fig2}).}

\textbf{Observation 2} indicates that if we select a relatively small number of antennas for beam training, the users can be regarded as residing in the far-field region of the sub-array.

\begin{figure*}[t]
	\centering
	\includegraphics[width=1\textwidth]{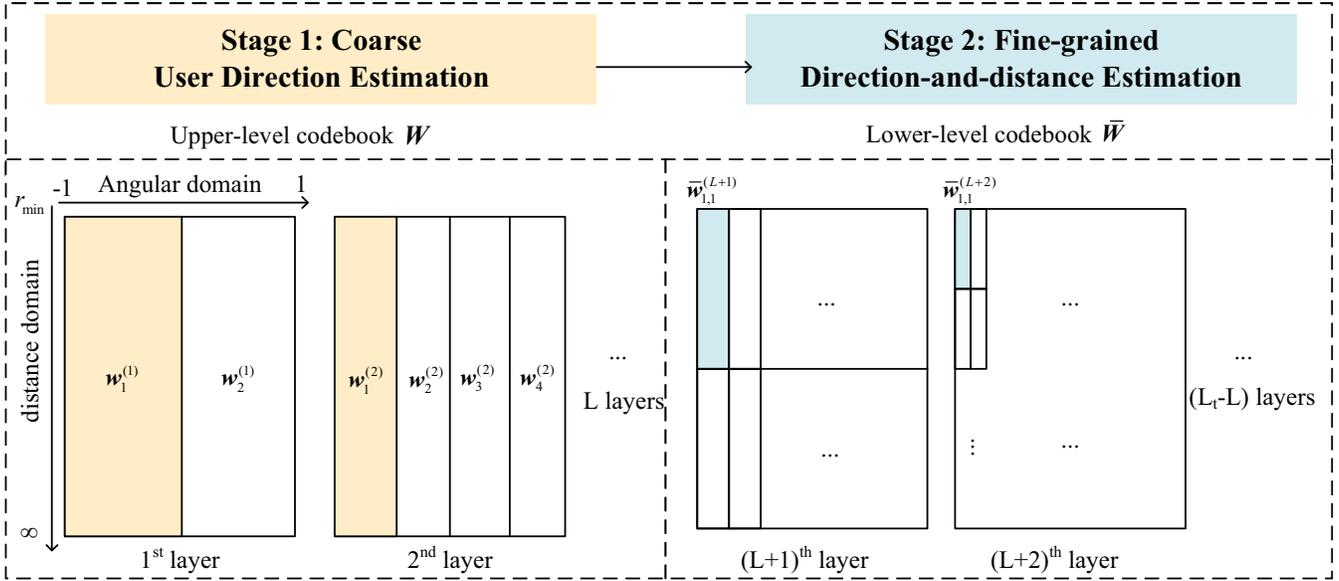}\\
	\caption{Illustration of the proposed two-stage hierarchical beam training method (codebook).}\label{fig3}
\end{figure*}

Motivated by the above, we propose a new \textit{two-stage hierarchical beam training} method, as illustrated in Fig.~\ref{fig3}. Specifically, in the first stage, we estimate the \textit{coarse user direction} by employing the sub-array for implementing the conventional far-field hierarchical beam training based on an upper-level codebook in the angular domain. Then, in the second stage, we further resolve \textit{the user direction as well as its effective distance} by designing a dedicated lower-level codebook in both the angular and distance domains as well as devising a new beam training method. Referring to the far-field binary-tree based hierarchical beam training~\cite{xiao}, the total number of layers of our hierarchical codebook can be set as $L_t=\log_2(N)$, while the upper-level and lower-level codebooks include $L$ and $L_t-L$ layers, respectively. The details will be presented in subsequent subsections.


\subsection{Stage 1: Coarse User Direction Estimation}

\subsubsection{Upper-level 1D Hierarchical Codebook Design}

For Stage 1, we first devise its upper-level one-dimensional (1D) hierarchical codebook in the angular domain and then present the beam training method for coarse user direction estimation. 

Specifically, similar to the conventional far-field hierarchical codebook, the proposed upper-level codebook includes $L$ layers of codewords based on the following two criteria. Firstly, for each layer in the upper-level codebook, each codeword should cover a specific angular region as well as the entire distance domain, i.e., $[r_{\min},\infty]$. Moreover, all the codewords in each layer should equally divide the entire spatial angular domain $[-1,1]$. Secondly, the beam coverage of an arbitrary codeword within a layer should be covered by the union of those of two codewords in the next layer.

To satisfy the above two criteria, the upper-level codebook is devised as follows.  For the $\ell$-th layer, $N_\ell=2^\ell$ antennas in the center of the BS XL-array are activated, which can generate a far-field beam with beam width of $2/N_\ell$~\cite{bisection}. Note that with a smaller $N_{\ell}$, the user is more likely to reside in the far-field region of the BS. 
As such, for the $\ell$-th layer, it has $N_\ell$ codewords, each pointing to a spatial direction of  $-1+\frac{2i-1}{N_\ell},\ell=1,...,N_\ell$.
Let $\mathbf{W}^{(\ell)}=\{\mathbf{w}_1^{(\ell)},...,\mathbf{w}_{N_\ell}^{(\ell)}\}$ denote the set of the $\ell$-th layer codewords in the first stage. Mathematically, the $i$-th codeword in $\mathbf{W}^{(\ell)}$ is given by


\begin{equation}
	\begin{aligned}\label{codeword_lower}
		\mathbf{w}_i^{(\ell)}=\left[ \mathbf{0}^T_{\frac{N-N_\ell}{2}},\mathbf{b}^T\left( -1+\frac{2i-1}{N_\ell},\infty\right), \mathbf{0}^T_{\frac{N-N_\ell}{2}}\right]^T ,\\ i=1,...,N_\ell.
	\end{aligned}
\end{equation}
It is worth noting that $\mathbf{b}(\theta,\infty)$ is equal to $\mathbf{a}(\theta)$. 

\subsubsection{Hierarchical Beam Training of Stage 1}\label{4.2.2}

Based on the designed upper-level codebook, the angular-domain hierarchical beam training in Stage 1 can be implemented by using the binary-tree based beam search. Suppose that $\mathbf{w}^{(L)}_{i^*}$ is the optimal codeword in the $L$-th layer of Stage 1 with $i^*$ denoting the best codeword index. The coarse direction is then estimated as $\theta^*=-1+\frac{2i^*-1}{2^L}$, while the true direction lies in the angular region $[-1+\frac{2i^*-2}{2^L},-1+\frac{2i^*}{2^L}]$.  

For Stage 1, it takes two training beams for each layer of beam training and thus the total overhead is $2L$, which is consistent with the conventional far-field case \cite{bisection}. 

\vspace{0.4em}
\noindent\underline{\textbf{Remark 1.}} \textit{(Deactivation approach)
Note that in the conventional far-field hierarchical codebook design \cite{bisection}, the activated sub-array can be selected arbitrarily from the entire array. In contrast, for the proposed first-stage codebook design, the activated sub-array are selected as those located in the middle of the full-array for improving the beam training performance. This is because the use of non-central sub-array may cause a larger estimation error as it may not point to the desired spatial directions during beam training. For the same reason, the joint sub-array and deactivation technique proposed in \cite{xiao} is not applicable to our design.  }

\subsection{Stage 2: Fine-grained Direction-and-distance Estimation}
Given the coarse user direction estimated in Stage 1, the optimal near-field beam vector for the user can be found by exhaustively searching over all candidate direction-and-distance pairs. However, the beam training overhead of this method is the product of the number of remaining candidate directions, i.e., $2^{L_t-L}$, and the number of sampled distances $S$, and hence requiring $2^{L_t-L}S$ training overhead in the second stage. To further reduce the overhead, we devise below a new lower-level \textit{two-dimensional (2D)} hierarchical codebook as well as its dedicated beam training design.

\subsubsection{Lower-Level 2D Hierarchical Codebook Design}

As illustrated in Fig.~\ref{fig3}, the lower-level codebook in Stage 2 consists of $L_t-L$ layers of beam codewords, which is designed to satisfy the following criteria to facilitate the binary-tree search in the hierarchical beam training (see Section \ref{4.3.2}). First, each codeword of the lower-level codebook should cover a specific angular region as well as a distance region in the polar domain. Moreover, all the codewords in each layer should jointly cover the entire angular domain $[-1,1]$ and the distance domain $[r_{\min},\infty]$. Second, the beam coverage of an arbitrary codeword in one layer should be covered by the union of those of two codewords in the next layer.



To satisfy the above criteria, we design the lower-level codebook as follows. 
For the $u$-th layer, $u=L+1,...,L_t$, $N_u=2^u$ antennas in the center of the BS XL-array are activated. Denote the set of the $u$-th layer codewords as $\bar{\mathbf{W}}^{(u)}=\{\bar{\mathbf{w}}_{1,1}^{(u)},...,\bar{\mathbf{w}}_{1,S_u}^{(u)},...,\bar{\mathbf{w}}^{(u)}_{i,1},...,\bar{\mathbf{w}}_{i,S_u}^{(u)},...\}$, 
wherein each codeword $\bar{\mathbf{w}}^{(u)}_{i,j}, i=1,...,N_u,j=1,...,S_u$, corresponds to a near-field steering vector pointing to a specific direction-and-distance pair, i.e., $\mathbf{b}(\theta_i^{(u)},r_{i,j}^{(u)})$, and $S_u=2^{u-L}$ is the number of sampled distances for $\theta_i^{(u)}$ in the $u$-th layer. The design of codeword $\bar{\mathbf{w}}^{(u)}_{i,j}$ is specified as follows:

Firstly, for the $u$-th layer, each codeword points to a spatial direction such that the sampled directions equally divide the angular domain $[-1,1]$, i.e., $\theta_i^{(u)}=-1+\frac{2i-1}{N_u},i=1,...,N_u$. Mathematically, we have

\begin{figure}[t]
	\centering
	\includegraphics[width=0.45\textwidth]{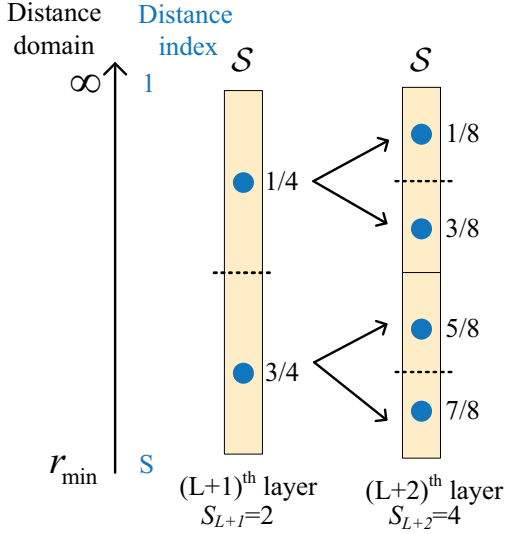}\\
	\caption{Illustration of the sampled distances of the lower-level codebook.}\label{dis_codebok}
\end{figure}

\begin{equation}
	\begin{aligned}\label{codeword_upper}
		\bar{\mathbf{w}}_{i,j}^{(u)}=\left[ \mathbf{0}^T_{\frac{N-N_u}{2}},\mathbf{b}^T\left( \theta_i^{(u)},r_{i,j}^{(u)}\right), \mathbf{0}^T_{\frac{N-N_u}{2}}\right]^T, \\ i=1,...,N_u,j=1,...,S_u.
	\end{aligned}
\end{equation}

Next, consider the sampled distances for the codewords with the same direction $\theta_i^{(u)}$. To cover the distance domain sampled in \eqref{5}, we define a \textit{distance index set} as 
\begin{equation}\label{dis_codebook}
	\mathcal{S}\triangleq\left\lbrace 0,...,S-1\right\rbrace.
\end{equation}
Then, for each spatial direction $\theta_i^{(u)}$ of the $u$-th layer codewords, 
$S_u$ sampled distances are chosen such that the distance index set are progressively divided in the binary-tree manner, as illustrated in Fig.~\ref{dis_codebok}. Specifically, let $I(u,j)$ denote a function that gives the index of the $j$-th sampled distance of the $u$-th layer codewords, which is given by 
\begin{equation}
	I(u,j)=\left\lceil\frac{2j-1}{2S_u}S\right\rceil.
\end{equation}
Consequently, the steering sampled distance of codeword $\bar{\mathbf{w}}_{i,j}^{(u)}$ is

\begin{equation}
	r^{(u)}_{i,j}=\frac{S_\Delta}{\mathcal{S}[I(u,j)]}(1-(\theta_i^{(u)})^2).
\end{equation}

\begin{algorithm}[!t]\label{method2}
	\caption{Proposed two-stage hierarchical beam-training}
	\begin{algorithmic}[1]
		\STATE {\textbf{Input:} $N,f,L,S_\Delta,S$.}
		\STATE \textbf{{Stage 1: Coarse user direction estimation}}
		\STATE \quad \textbf{for} $\ell=1,...,L$ \textbf{do} 
		\STATE \quad \quad Calculate the upper-level hierarchical codebook according to \eqref{codeword_lower}.
		\STATE \quad \quad Test two codewords and find the best codeword index by measuring the received signal power of \eqref{rpower}.
		\STATE \quad \textbf{end for} 
		\STATE \textbf{{Stage 2: Fine-grained direction-and-distance estimation}}
		\STATE \quad \textbf{for} $u=L+1,...,L_t-1$ \textbf{do} 
		\STATE \quad \quad Calculate the lower-level hierarchical codebook according to \eqref{codeword_upper}.
		\STATE \quad \quad Test four codewords and find the best codeword index by measuring the received signal power of \eqref{rpower}.
		\STATE \quad \textbf{end for} 
		\STATE \quad Calculate the sampled distances of the last layer by \eqref{last_layer}.
		\STATE \quad Test the signal power of the last-layer codewords.
		\STATE {\textbf{Output:} The best codeword $\mathbf{w}^*$, and the estimated user direction and distance.}
	\end{algorithmic}
\end{algorithm}

\subsubsection{Hierarchical Beam Training of Stage 2}\label{4.3.2}

Based on the devised lower-level codebook, the second-stage beam training design is detailed as the following two cases: 

1) \textbf{Layers} $u=L+1,...,L_t-1$. For the lower layers except the last layer, we implement
the binary-tree based beam search over both the angular and the distance domain to gradually resolve the fine-grained direction-and-distance of the user. Specifically, for each layer of beam training, four codewords in \eqref{codeword_upper} are tested, corresponding to the combinations of two steering directions and two steering distances. According to the received signal power at the user, we can decide the best beam in the $u$-th layer as well as its steering direction and distance. Then, we enter the next layer and repeat this procedure.  

2) \textbf{Last layer}, i.e., $u=L_t$. For the last layer, we aim to determine the best near-field beam for the user to maximize the beam training accuracy. Let $\bar{\mathbf{w}}_{i^*,j^*}^{(L_t-1)}$ denote the best codeword obtained in layer $(L_t-1)$ with $(i^*,j^*)$ denoting the corresponding beam index. Then, for the beam training in Layer $L_t$, the two beam directions are selected as  $\theta_1=-1+\frac{2(2i^*-1)-1}{N}$ and $\theta_2=-1+\frac{2(2i^*)-1}{N}$. For each direction, we select all the beam codewords in $\mathcal{S}$ whose steering distance indices are within $[I(L_t-1,j^*-1),I(L_t-1,j^*+1)]$. Mathematically, the steering distances of the selected codewords for direction $\theta_i$ with $i=1,2$ are
\begin{equation}\label{last_layer}
	r\in \left\lbrace \frac{S_\Delta}{\mathcal{S}[k]}(1-(\theta_i)^2)\right\rbrace ,
\end{equation}
where $I(L_t-1,j^*-1)<k<I(L_t-1,j^*+1)$.

Accordingly, for Stage 2, it requires testing $4$ codewords for each layer $u=L+1,...,L_t-1$ and the remaining $\lceil \frac{S}{2^{L_t-L}} \rceil$ codewords for the last layer $u=L_t$. The total beam training overhead of Stage 2 is $4(L_t-L-1)+\lceil \frac{S}{2^{L_t-L}} \rceil$. The detailed procedure of the proposed two-stage hierarchical beam training design is summarized in \textbf{Algorithm 1}.

An example is provided as below to demonstrate the detailed procedure of the proposed two-stage hierarchical beam training method.

\vspace{0.4em}
\noindent\underline{\textbf{Example 1:}} \textit{Consider the case with $N=512$. The user is located at $[\theta,r]=[0,40\,m]$. The parameters are set as $L_t=9$, $L=7$, and $S_\Delta=70$. As introduced in Section \ref{4.2.2}, the coarse user direction is gradually estimated as $-0.0078$ after 7 layers of beam training in Stage 1 (the best codeword of Stage 1 is $\mathbf{w}_{64}^{(7)}$). Then, we enter the second stage. Four codewords are tested in the 8-th layer, i.e., $\bar{\mathbf{w}}^{(8)}_{i,j},i=127,128,j=1,2$, with their sampled distances given by $r^{(8)}_{i,1}=68(1-(\theta_i^{(8)})^2)$, $r^{(8)}_{i,2}=17(1-(\theta_i^{(8)})^2)$. Suppose that $\bar{\mathbf{w}}^{(8)}_{128,1}$ is the best codeword for the 8-th layer. Then, in the last layer, six beam codewords are tested, i.e., $\bar{\mathbf{w}}^{(9)}_{i,j},i=256,257,j=1,2,3$, corresponding to the sampled distances  $r\in[\infty,68,34]$ (see \eqref{last_layer}). Finally, if $\bar{\mathbf{w}}_{256,3}^{(9)}$ yields the maximum received signal power at the user, the best codeword of the proposed two-stage hierarchical beam training method is identified as $\bar{\mathbf{w}}^{(9)}_{256,3}$, which steers to the location 
	$[\theta,r]=[-0.0019, 35\,m]$. }
	
\section{Discussions and Extensions}
In this section, we discuss several key parameters of our proposed beam training design as well as its extensions to other systems with planar array and metasurface. 
\subsection{Beam Training Overhead} Based on the above introduction, the total training overhead of the proposed two-stage hierarchical method is given by 
\begin{equation}\label{overhead}
	\begin{aligned}
		T&=2L+4(L_t-L-1)+\left\lceil \frac{S}{2^{L_t-L}}\right \rceil
		\\ &\overset{(a)}{\approx} 2L+4(L_t-L)
		\\&=2\log_2(N_L)+4\log_2(\frac{N}{N_L})\sim\mathcal{O}(\log_2(N)),
	\end{aligned}
\end{equation}
where $(a)$ is obtained by assuming that four codewords are tested in the last layer, similar as the preceding layers of Stage 2. Generally, it takes 2 training beams for each layer of Stage 1 for coarse user direction estimation, followed by 4 training beams for each layer of Stage 2 to fine-tune both the beam direction and distance. Note that the approximation in \eqref{overhead} is not affected by $S$ since $N$ is practically much larger than $S$. According to \eqref{overhead}, the training overhead of our design scales in the order of $\mathcal{O}(\log_2(N))$, which is significantly lower than the benchmark  near-field beam training schemes introduced in Section \ref{sec3}.

\vspace{0.4em}

\noindent\underline{\textbf{Example 2:}} \textit{Consider the same system setup as that in \textbf{Example 1} with random user locations. The total beam training overhead of our proposed design is  $2\times7+4+6=24$.}

%

\begin{figure}[!htb]
	\centering
	\subfigure[Normalized beam gain of the central sub-array with $N_L=N/4$ antennas ($L=L_t-2$).]{\includegraphics[width=0.5\textwidth]{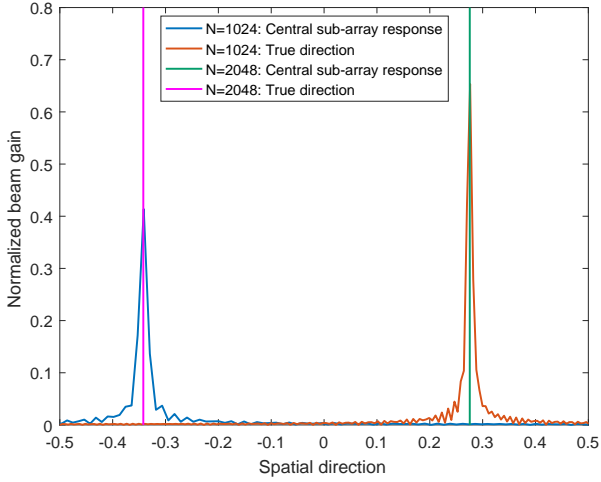}\label{subarray2}}\hspace{1cm}
	\subfigure[Normalized beam gain for the cases of $L=L_t-2$ and $L=L_t-3$ with $N=512$ and the BS-user distance $r=5$ m.]{\includegraphics[width=0.5\textwidth]{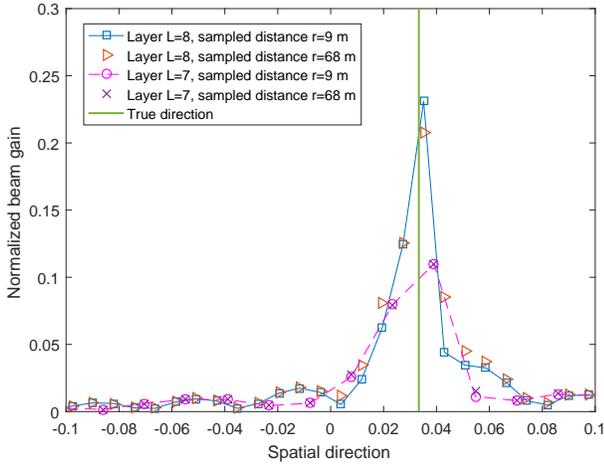}\label{proper}}
	\caption{Numerical results for the proper choice of $L$.}
	\label{fig: figure}
\end{figure}

\subsection{Proper Value of $L$}

In fact, there exists a trade-off between the beam training overhead and its accuracy with different $L$. In particular, when $L=L_t$, the proposed hierarchical codebook reduces to the conventional far-field hierarchical codebook \cite{bisection}. Generally, for minimizing the beam training overhead, it is desirable to choose $L$ as large as possible. However, for guaranteeing the beam training performance in the near field, we should properly choose a small number of activated antennas in Stage 1 (a small $L$) to avoid the \textit{energy-spread} effect, which is affected by the number of activated antennas and user distance. To balance the above trade-off, we first show in Fig. \ref{subarray2} that if the user is located in the Fresnel region ($r\geq r_{\min}=\frac{1}{2}\sqrt{\frac{D^3}{\lambda_c}}=\frac{\lambda_c}{4}\sqrt{\frac{N^3}{2}}$), there is little energy-spread during the first-stage beam training when $L\leq L_t-2$. Then, we plot in Fig. \ref{proper} the beam gain of all the codewords in \eqref{codeword_upper} for $L=L_t-2$ and $L=L_t-3$, respectively. It can be observed that the beam gain of the codewords with the first sampled distance ($r=9$ m) approximates that of the codewords with the second sampled distances  ($r=68$ m) for $N=512, L=7$. This indicates that if Stage 2 starts at an earlier layer than $L=L_t-2$, there is little enhancement in the spatial resolution for our proposed two-stage hierarchical beam training method, yet with increased training overhead as given in \eqref{overhead}. Therefore, $L$ can be practically chosen as $L_t-2$ for an arbitrary number of antennas in the XL-array without loss of generality. On the other hand, if the user locates in a region with the BS-user distance larger than $R_c$ (not equal to $r_{\min}$ but known as a prior), the maximum number of activated antennas in Stage 1, $N_L$, should be chosen such that the user locates within the Fresnel boundary of $N_L$, i.e., $N_L=\lfloor (\frac{4R_c}{\lambda_c})^{\frac{2}{3}} \rfloor$. Correspondingly, the number of layers in Stage 1 should be $L=\lceil \log_2(N_L)\rceil$.

\begin{figure}[t]
	\centering
	\includegraphics[width=0.45\textwidth]{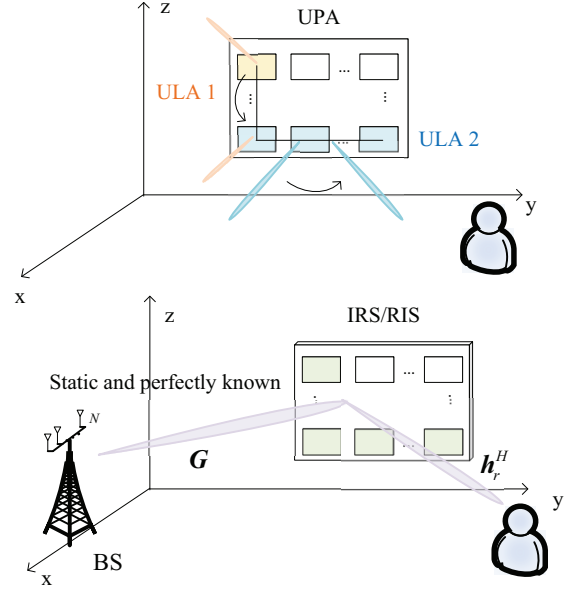}\\
	\caption{Extension to UPA and IRS.}\label{fig4}
\end{figure}

\subsection{Extensions to UPA and IRS:}

 \textit{1) UPA:} 
	The proposed two-stage hierarchical beam training method for ULA can be readily extended to the systems with a uniform planar array (UPA) at the BS. 
	
	Without loss of generality, we assume that the UPA is composed of $M=M_y\times M_z$ elements placed in the $y$-$z$ plane, and the user is located in the $x$-$y$ plane, as shown in Fig. \ref{fig4}. Denote $(0,y_0,z_0)$ and $(x_u,y_u,0)$ as the coordinate of the (1,1)-th element at the left-bottom of the UPA and the coordinate of the user, respectively. Then, the coordinate of the $(m_y,m_z)$-th element of the UPA can be represented as $(0, y_{m_y}, z_{m_z})$ with $y_{m_y} =
	y_0+(m_y-1)d$, $z_{m_z} = z_0+(m_z-1)d$, $m_y = 1,...,M_y$, $m_z =
	1,...,M_z$. Besides, the distance from the $(m_y,m_z)$-th element to the user is $D_r(m_y,m_z)=\sqrt{(x_{u})^2+(y_{m_y}-y_u)^2+(z_{m_z})^2}$. Based on the
	spherical wavefront propagation model, the near-field LoS channel between the UPA and the user can be represented by
	\begin{equation}\label{hr}
\mathbf{h}^H_r=\alpha\mathbf{c}^H_r,
	\end{equation}
where $\alpha=\frac{\sqrt{{\rho_0}}}{D_r(1,1)}$ is the complex channel gain between the IRS and user, and $\mathbf{c}_r$ is the near-field steering vector, given by
	\begin{equation}\label{18}
	\begin{aligned}
		\mathbf{c}_r=\Big[ e^{-\jmath{\frac{2\pi}{\lambda_c}}D_r(1,1)}, ..., e^{-\jmath{\frac{2\pi}{\lambda_c}}D_r(1,M_z)},...,\\
		e^{-\jmath{\frac{2\pi}{\lambda_c}}D_r(M_y,1)},...,e^{-\jmath{\frac{2\pi}{\lambda_c}}D_r(M_y,M_z)}				
		\Big]^T. 
	\end{aligned}
\end{equation}

Based on the UPA-user channel in \eqref{hr}, the received signal at the user is given by
\begin{equation}\label{r_upa}
	y(\mathbf{v})=\mathbf{h}_r^H\mathbf{v}x+z_0.
\end{equation}

To facilitate beam training for the user, we can decouple the joint
three-dimensional (3D) XL-array beam training into two
sequential phases, corresponding to the horizontal and
vertical beam training for two mutually perpendicular
ULAs, see e.g., ULA 1 and ULA 2 in Fig. 6.
As such, our proposed method can be adopted directly to the UPA case.
	
	
	\textit{2) IRS/RIS:} Intelligent reflecting surface (IRS) \cite{zr_survey}, also known as reconfigurable intelligent surface (RIS) \cite{ris_survey2,survey_ris}, has emerged as a promising technology for enhancing the spectral efficiency of future wireless systems, by leveraging a massive number of low-cost passive reflecting elements to reconfigure the wireless propagation environment. In practice, the large-size IRS is usually deployed near the users to enhance their communication performance with associated BSs. This renders the users more likely to reside in the near field of the IRS, and hence making the beam training for the cascaded IRS channel more challenging. 
	
	First, the IRS-user channel can be modeled similarly as \eqref{hr} and hence is omitted here. Assume that the BS is equipped with a ULA placed along the $x$-axis and the IRS is located in the far-field region of the BS. Then, the LoS channel between the BS and IRS can be modeled as\cite{zr_survey}
	\begin{equation}\label{G}
		\mathbf{G}=\beta\mathbf{c}_t\mathbf{a}^H(\theta_t),
	\end{equation}
	where $\beta=\frac{\sqrt{{\rho_0}}}{D_t(1,1)}$ is the complex channel gain between the BS and IRS, $\theta_t$ is the spatial AoD from the BS antennas to the IRS, $\mathbf{c}_t$ is the near-field steering vector, given by 
	\begin{equation}
		\mathbf{c}_t=\big[ e^{-\jmath{\frac{2\pi}{\lambda_c}}D_t(1,1)}, ...,e^{-\jmath{\frac{2\pi}{\lambda_c}}D_t(M_y,M_z)}				
		\big]^T,
	\end{equation}
	 where $D_t(m_y,m_z)=\sqrt{(x_{b})^2+(y_{m_y})^2+(z_{m_z}-z_b)^2},m_y = 1,...,M_y, m_z =
	1,...,M_z$ denotes the distance from the $(m_y,m_z)$-th IRS element to the BS, $(x_b,0,z_b)$ is the coordinate of the BS. Let $\boldsymbol{\Phi}\triangleq{\rm diag}(e^{\jmath\phi_1},e^{\jmath\phi_2},...,e^{\jmath\phi_M})$ denote the diagonal IRS reflecting matrix, where for simplicity we assume
	that the reflection amplitude of each element is set to one, and $\phi_m,m=1,...,M$ denotes the phase-shift of the $m$-th element. Based on the channel models in \eqref{hr} and \eqref{G}, the received signal at the user is given by
	
	\begin{equation}\label{r_signal}
			y(\mathbf{v},\boldsymbol{\Phi})=\mathbf{h}_r^H\boldsymbol{\Phi}\mathbf{G}\mathbf{v}x+z_0.
	\end{equation}

\begin{table*}[]
	\caption{Comparison of Beam Training Overhead of Different  Schemes}\label{table}
	\centering
	\begin{tabular}{|l|l|l|}
		\hline
		\textbf{Beam training method}                         & \textbf{Total overhead} & \textbf{In our setup}\\ \hline
		Exhaustive-search based near-field beam training  & $NS$       & 3072     \\ \hline
		Two-phase near-field beam training   & $N+S$ &518   \\ \hline	
		Far-field exhaustive-search based beam training         & $N$    &512  \\ \hline
		Far-field hierarchical beam training   & $2\log_2(N)$    & 18    \\ \hline
		Proposed two-stage hierarchical beam training &  $2\log_2(N_L)+4\log_2(\frac{N}{N_L})$  &24     \\ \hline	
	\end{tabular}
\end{table*}

Since the BS and IRS are usually deployed at fixed locations with a LoS path, the BS-IRS link can be considered as quasi-static and perfectly known at the BS by using existing channel estimation methods, see e.g., \cite{ce_survey,discrete_jsac,star_ce,weiyi}. To facilitate beam training for the user, we first design the BS transmit beamforming to align with the BS-IRS channel, i.e., $\mathbf{v}=\mathbf{a}(\theta_t)$. As such, the BS can be treated as having an equivalent single antenna, and the received signal in \eqref{r_signal} can be written as 
	\begin{equation}\label{r_signal2}
		y(\boldsymbol{\Phi})=\alpha\beta\mathbf{c}_r^H\boldsymbol{\Phi}\mathbf{c}_tx+z_0.
	\end{equation}

	Next, to remove the effect of the near-field BS-IRS steering vector, i.e.,  $\mathbf{c}_t$, we rewrite the IRS phase-shift matrix as $\boldsymbol{\Phi}=\bar{\mathbf{\Phi}}{\rm diag}(\mathbf{c}^H_t)$, where $\bar{\boldsymbol{\Phi}}\triangleq{\rm diag}(e^{\jmath\bar{\phi}_1},e^{\jmath\bar{\phi}_2},...,e^{\jmath\bar{\phi}_M})$ is defined as a pseudo phase-shift matrix of the IRS used for modifying the directional beams during beam training, and ${\rm diag}(\mathbf{c}^H_t)$ is a constant phase-shift matrix reserved for counteracting $\mathbf{c}_t$. By doing so, \eqref{r_signal2} can be further written as
	\begin{equation}\label{r_signal3}
		y(\bar{\boldsymbol{\phi}})=\alpha\beta\mathbf{c}_r^H\bar{\boldsymbol{\phi}}x+z_0,
	\end{equation}
	where $\bar{\boldsymbol{\phi}}=[e^{\jmath\bar{\phi}_1},e^{\jmath\bar{\phi}_2},...,e^{\jmath\bar{\phi}_M}]^T$. 
We can easily observe that \eqref{r_signal3} has a similar form as \eqref{r_upa} by treating $\bar{\boldsymbol{\phi}}$ as the transmit beamforming vector $\mathbf{v}$. Hence, the beam traning for the cascaded IRS channel reduces to the near-field beam training for the IRS~\cite{csyou1},
	which is similar to the case of a UPA. Based on the above discussions, our proposed beam training algorithm can also be applied to IRS-aided near-field communication systems.

\section{Numerical Results}\label{sec5}
We provide numerical results in this section to validate the effectiveness of our proposed two-stage hierarchical beam training method. Consider an XL-array communication system with $N=512$ and $f=100$ Ghz. Hence, we have $\lambda_c=0.003$ m, $S_\Delta=68.27$, $S=6$, $r_{\min}=6.14$ m~\cite{nf_exhaustive}. The reference SNR is $\gamma=\frac{PN\beta_0}{r^2\sigma^2}$ with $\beta_0=-72$ dB, $P=30$ dBm. 
The noise power is set as $\sigma^2=-80$ dBm. To characterize the beam identification accurcy, we define $P_{\rm suc}=\mathbb{E}(\mathbb{I}((\hat{n}^*,\hat{s}^*)=(n^*,s^*)))$ as the success rate under the polar-domain codebook, where $(\hat{n}^*,\hat{s}^*)$ denotes the obtained codeword index by the specific beam training scheme and $(n^*,s^*)$ is the optimal codeword index in \eqref{opt_codeword}. The achievable rate in bits/second/Hz (bps/Hz) is given by $R=\log_2(1+\gamma|\mathbf{h}^H\mathbf{v}|^2)$. All the numerical results are averaged over 5000 random channel realizations. 

For performance comparison, the following benchmark schemes are considered:

\begin{figure}[t]
	\centering
	\includegraphics[width=0.5\textwidth]{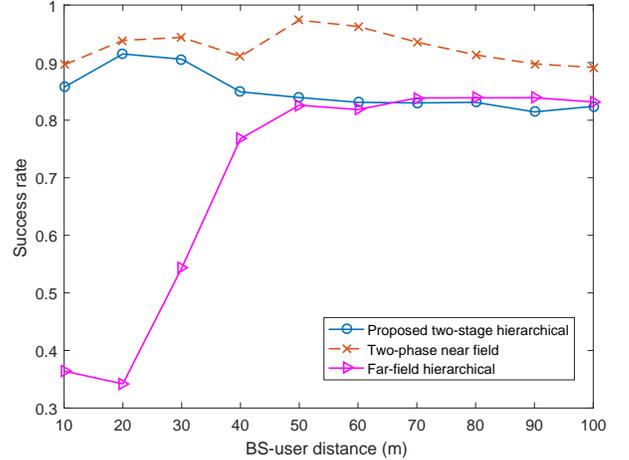}\\
	\caption{Success rate versus BS-user distance.}\label{suc_dis}
\end{figure}

\begin{figure}[t] 
	\centering
	\includegraphics[width=0.5\textwidth]{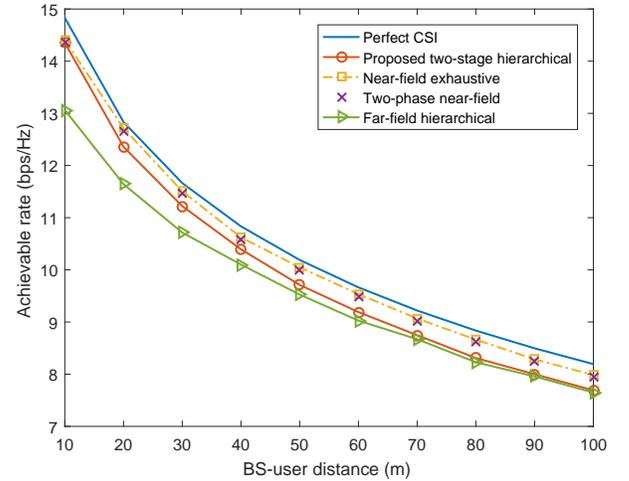}\\
	\caption{Achievable rate versus BS-user distance.}\label{rate_dis}
\end{figure}

\begin{itemize}
	\item \textbf{Perfect-CSI based beamforming}: In this scheme, the transmit beamforming vector $\mathbf{v}$ is designed to perfectly align with the user's channel, i.e., $\mathbf{v}^*=\mathbf{b}(\theta,r)$. This scheme serves as a performance upper-bound when evaluating the achievable rate of our proposed scheme.
	\item \textbf{Exhaustive-search based near-field beam training}: as introduced in Section \ref{3.1}.		
	\item \textbf{Two-phase near-field beam training} 
	with $K=1$~\cite{two_phase}. The detailed procedures are presented in Section~\ref{3.2}.
	\item \textbf{Conventional far-field hierarchical beam training} \cite{bisection}: In this scheme, the conventional far-field hierarchical codebook is adopted for beam training.	
\end{itemize}

\textit{1) Overhead Comparison:} We first compare the beam training overhead of different beam training schemes in Table \ref{table}. It is observed that the proposed two-stage hierarchical beam training scheme requires a much smaller number of training symbols than other near-field beam training schemes. Specifically, our proposed scheme can achieve over $99\%$ and $95\%$ overhead reduction as compared to the exhaustive search-based and two-phase beam training schemes, which, however, only suffers little rate performance loss as will be shown in Figs. \ref{rate_dis} and \ref{rate_snr}.

\begin{figure}[t]
	\centering
	\includegraphics[width=0.5\textwidth]{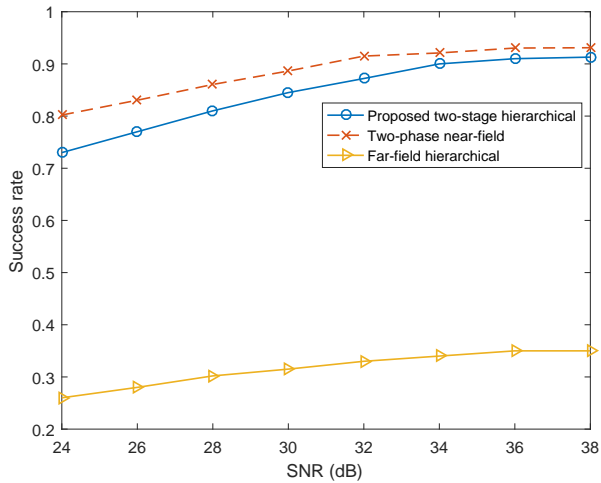}\\
	\caption{Success rate versus SNR.}\label{suc_snr}
\end{figure}

\begin{figure}[t]
	\centering
	\includegraphics[width=0.5\textwidth]{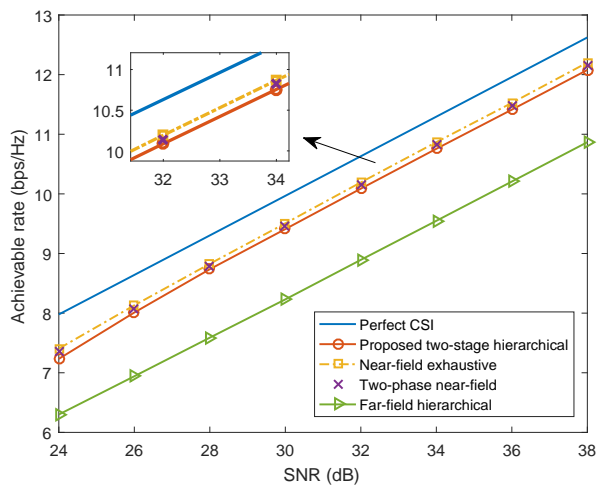}\\
	\caption{Achievable rate versus SNR.}\label{rate_snr}
\end{figure}

\textit{2) Impact of BS-user Distance:} Fig.~\ref{suc_dis} shows the effect of the BS-user distance on the success rate by different beam training schemes. It is observed that our proposed scheme with extremely small training overhead achieves comparable success rate with the two-phase scheme. Moreover, the proposed two-stage hierarchical beam training scheme maintains a relatively high success rate (i.e., higher than $80\%$) for all distances. This implies that our design is not sensitive to the BS-user distance, thus being applicable to both far-field and near-field communication scenarios. In contrast, the conventional far-field hierarchical beam training scheme suffers a significantly low success rate when the BS-user distance is short (e.g., $r<50$ m). This shows that the far-field hierarchical beam training scheme is not applicable to the near field since it neglects the spherical-wavefront channel characteristic.

Besides, Fig.~\ref{rate_dis} illustrates the rate performance of different beam training schemes versus the BS-user distance.  Several key observations are made as follows. Firstly, the proposed scheme significantly outperforms the far-field hierarchical beam training scheme, especially when the user is close to the XL-array. This attributes to our second-stage design, which effectively takes into account the near-field channel characteristics. Secondly, our proposed scheme attains close rate performance with the two-phase based near-field beam training scheme especially when the BS-user distance is small, yet with substantial reduced training overhead (i.e., 24 versus 518). Finally, the rate performance of our proposed scheme approximates that of the exhaustive-search based scheme when the BS-user distance is small, and finally degenerates to that of the far-field hierarchical beam training scheme when the BS-user distance is large. This also verifies that our proposed two-stage hierarchical beam training method is universal for both near-field and far-field communications.

\textit{3) Impact of the Effective SNR:}   Fig. \ref{suc_snr} shows the success rate of different beam training schemes versus the effective SNR. It is observed that the success rate of our proposed two-stage hierarchical beam training scheme monotonically increases with the SNR, and finally converges to that of the two-phase near-field beam training scheme. This is because the beam identification error caused by AWGN starts to be negligible in the high-SNR regime. In contrast, the success rate of conventional far-field hierarchical beam training scheme is always lower than $40\%$, even under a high SNR (over 30 dB). This is because the far-field hierarchical beam training scheme fails to capture the distance information of the user.

Fig. \ref{rate_snr} presents the performance of achievable rate versus the effective SNR under different beam training schemes. It is observed that our proposed scheme always outperforms the conventional far-field hierarchical beam training scheme by over 1 bps/Hz, yet without compromising much rate performance compared with the exhaustive-search based near-field beam training scheme.

\begin{figure}[t] 
	\centering
	\includegraphics[width=0.5\textwidth]{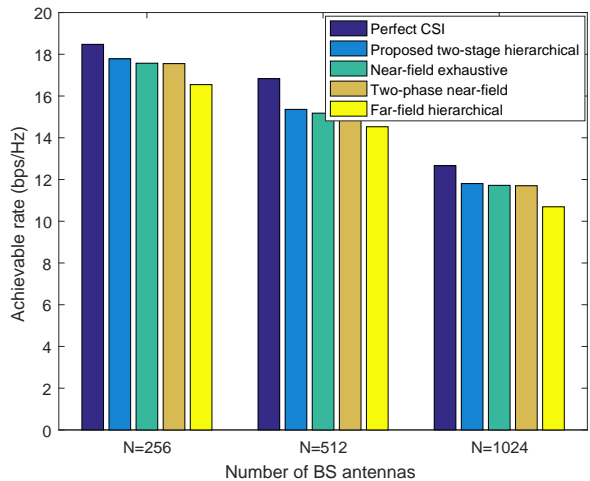}\\
	\caption{Achievable rate versus number of antennas, $N$.}\label{rate_N}
\end{figure}

\textit{4) Impact of the Number of BS Antennas:} Finally, in Fig.~\ref{rate_N}, we plot the achievable rates of different schemes under different number of BS antennas $N$. The users are assumed to be located near the Fresnel boundary. It is observed that our proposed scheme always outperforms the conventional far-field counterparts. 
This again validates that our design is robust under different system setups.

\section{Conclusion}\label{sec6}
In this paper, we proposed a novel two-stage hierarchical beam
training method for near-field communications, where a coarse user direction is found in the first stage, followed by a second stage that further resolves fine-grained direction-and-distance of the user. We proposed efficient hierarchical codebook design for each stage as well as its corresponding beam training method. It was shown that our proposed method significantly reduces the training overhead of the existing near-field beam training methods, yet achieving comparable rate performance.



\bibliographystyle{IEEEtran}
\bibliography{mybib}

\end{document}